\documentstyle [12pt,a4p,epsfig,amsmath,multicol]{article}
\begin{document}
\vspace*{0.6cm}

\begin{center} 
{\normalsize\bf Comment on `Measuring propagation speed of Coulomb fields'
     [R. de Sangro {\it et al.} Eur. Phys. J.C (2015) 75:137]}
\end{center}
\vspace*{0.6cm}
\centerline{\footnotesize J.H.Field}
\baselineskip=13pt
\centerline{\footnotesize\it D\'{e}partement de Physique Nucl\'{e}aire et 
 Corpusculaire, Universit\'{e} de Gen\`{e}ve}
\baselineskip=12pt
\centerline{\footnotesize\it 24, quai Ernest-Ansermet CH-1211Gen\`{e}ve 4. }
\centerline{\footnotesize E-mail: john.field@cern.ch}
\baselineskip=13pt
 
\vspace*{0.9cm}
\abstract{ Some remarks are made on the theoretical interpretation of
 recent experimental results on the propagation speed of
 electromagnetic force fields and of previous related experimental and
 theoretical work}
 
\vspace*{0.9cm}
\vspace*{0.9cm}
\normalsize\baselineskip=15pt
\setcounter{footnote}{0}
\renewcommand{\thefootnote}{\alph{footnote}}

 \par The conclusion of the experiment described in Ref.~\cite{PSCF} is 
      found in the last sentence of the abstract of the paper:
   \begin{itemize}
     \item[] `The results we obtain, on a finite lifetime
            kinematical state are compatible with an electric field 
          rigidly carried by the beam itself'
      \end{itemize}
   This is another way of stating that the results are consistent with a propagation
   speed of Coulomb forces that is instantaneous, or at any rate much greater than
    the speed of light and not consistent with the retarded field of the Li\'{e}nard-Wiechert
    formula. Similar behaviour has been recently reported for magnetic force fields
    in ~\cite{KMSR}.
     \par This comment points out that such behaviour is predicted  by
      the Relativistic Classical Electro-Dynamics (RCED) theory developed by
    the present author~\cite{JHFRCED,JHFPPNC}. In this theory electric and magnetic
     force fields are mediated by the exchange of space-like virtual photons as in
      Quantum Electrodynamics. The equations of RCED are derived by considering 
      the classical limit of the QED invariant amplitude for M\o ller scattering:
      $\mathrm{ee}\rightarrow \mathrm{ee}$. Some comparisons of the theory with
     text-book Classical Electromagnetism
     have been worked out in~\cite{JHFIJMP}, including a demonstration, similar to
    that of Laplace for gravitational forces~\cite{Laplace}, of the impossiblity
    of stable circular Keplerian orbits with retarded electromagnetic force fields.
    See also the related work on the speed of gravitational forces reported in Ref.~\cite{TVF}.
    \par Earlier evidence for a superluminal propagation of microwave fields in the
     near zone was obtained by Mugnai, Ranfagni and Ruggeri~\cite{MRR}. Evidence for superluminal
     signal propagation in the near field region has also been obtained in some
     recent amateur experiments~\cite{AEE1,AEE2,AVR}.
     Indeed, as pointed out by Smirnov-Rueda~\cite{SRFP}, evidence for similar effects
     was even observed (but not published~\cite{Buchwald}) in Hertz' original experiments on
     electromagnetic wave propagation~\cite{Hertz}.
     \par Ref.~\cite{PSCF} cites a paper by the present author~\cite{JHFLW} where the
      derivation of retarded fields as calculated by Li\'{e}nard\cite{Lienard} and
     Wiechert~\cite{Wiechert}
      was critically reviewed. This showed, as previously pointed out by
     Whitney~\cite{Whitney}, that an elementary mathematical error due to
     miscalculation of the effective charge density of a moving charge occurs
     in the derivation of the retarded Li\'{e}nard-Wiechert potentials.
     Corrected formulas for retarded potentials and fields were given (see also
     Ref.~\cite{JHFBCFC}). However, this work is not related to the force
     fields measured in  Ref.~\cite{PSCF} which are instead consistent with the
     instantaneous RCED force fields mentioned above.
      \par In conclusion, the additional information, related to the work presented
       in Ref.~\cite{PSCF}, given in this comment, is:
        \begin{itemize}
         \item A theoretical prediction (Refs.~\cite{JHFRCED,JHFPPNC}) of the observed
                 non-retarded nature of Coulombic force fields.
         \item A precise reference~\cite{Laplace} to the work of Laplace demonstrating the
                 non-retarded nature of gravitational force fields.
          \item Citation of previous experimental results, consistent with the findings
                  of Ref.~\cite{PSCF}, but not  mentioned in it (Refs.~[8-13]).
          \item Pointing out that the paper~\cite{JHFLW} (cited as Ref.~[6] in ~\cite{PSCF}) 
                  which discusses only retarded force fields, has no direct 
                  relevance to the explanation of the results of ~\cite{PSCF}.

           \item Noting (as previously shown in Ref.~\cite{Whitney}), that the
                Li\'{e}nard-Wiechert potentials, used for calculation of retarded fields in
            Ref.~\cite{PSCF}, are invalid due to a simple mathematical error in their
             derivation, and that the corrected retarded potentials may be found in Ref.~\cite{JHFLW}.
               \end{itemize}
      \par {\bf Acknowledgements} \newline
      I would like to thank Guido Pizzella for informing me of the results presented in
      Ref.~\cite{PSCF} and Brendan Rycroft for pointing out to me the work of the 
      authors of Refs.~\cite{AEE1,AEE2,AVR}.  


\end{document}